\documentclass[aps,pre,10pt,twocolumn,a4paper,superscriptaddress]{revtex4}

\usepackage{graphicx}
\usepackage{amsmath, amssymb, bbm}
\usepackage[english]{babel}
\usepackage{color}
\usepackage[normalem]{ulem}

\begin{document}

\newcommand{\be}{\begin{equation}}
\newcommand{\ee}{\end{equation}}
\newcommand{\bq}{\begin{eqnarray}}
\newcommand{\eq}{\end{eqnarray}}
\newcommand{\bsq}{\begin{subequations}}
\newcommand{\esq}{\end{subequations}}
\newcommand{\bc}{\begin{center}}
\newcommand{\ec}{\end{center}}
\newcommand {\R}{{\mathcal R}}
\newcommand{\al}{\alpha}
\newcommand\lsim{\mathrel{\rlap{\lower4pt\hbox{\hskip1pt$\sim$}}
    \raise1pt\hbox{$<$}}}
\newcommand\gsim{\mathrel{\rlap{\lower4pt\hbox{\hskip1pt$\sim$}}
    \raise1pt\hbox{$>$}}}

\title{Junctions and spiral patterns in generalized Rock-Paper-Scissors models}

\author{P.P. Avelino}
\affiliation{Centro de Astrof\'{\i}sica da Universidade do Porto, Rua das Estrelas, 4150-762 Porto, Portugal}
\affiliation{Departamento de F\'{\i}sica e Astronomia, Faculdade de Ci\^encias, Universidade do Porto, Rua do Campo Alegre 687, 4169-007 Porto, Portugal}

\author{D. Bazeia}
\affiliation{Departamento de F\'{\i}sica, Universidade Federal da Para\'{\i}ba 58051-970 Jo\~ao Pessoa, Para\'{\i}ba, Brazil}

\author{L. Losano}
\affiliation{Departamento de F\'{\i}sica e Astronomia, Faculdade de Ci\^encias, Universidade do Porto, Rua do Campo Alegre 687, 4169-007 Porto, Portugal}
\affiliation{Centro de F\'{\i}sica do Porto, Rua do Campo Alegre 687, 4169-007 Porto, Portugal} 
\affiliation{Departamento de F\'{\i}sica, Universidade Federal da Para\'{\i}ba 58051-970 Jo\~ao Pessoa, Para\'{\i}ba, Brazil}

\author{J. Menezes}  
\affiliation{Departamento de F\'{\i}sica e Astronomia, Faculdade de Ci\^encias, Universidade do Porto, Rua do Campo Alegre 687, 4169-007 Porto, Portugal}
\affiliation{Centro de F\'{\i}sica do Porto, Rua do Campo Alegre 687, 4169-007 Porto, Portugal} 
\affiliation{Escola de Ci\^encias e Tecnologia, Universidade Federal do Rio Grande do Norte\\Caixa Postal 1524, 59072-970, Natal, RN, Brazil}

\author{B. F. Oliveira}  
\affiliation{Departamento de F\'{\i}sica, Universidade Federal da Para\'{\i}ba 58051-970 Jo\~ao Pessoa, Para\'{\i}ba, Brazil}


\date{\today}

\begin{abstract}
We investigate the population dynamics in generalized Rock-Paper-Scissors models with an arbitrary number of species $N$. We show, for the first time, that spiral patterns with $N$-arms may develop both for odd and even $N$, in particular in models where a bidirectional predation interaction of equal strength between all species is modified to include one N-cyclic predator-prey rule. While the former case gives rise to an interface network with Y-type junctions obeying the scaling law $L \propto t^{1/2}$, where $L$ is the characteristic length of the network and $t$ is the time, the later can lead to a population network with $N$-armed spiral patterns, having a roughly constant characteristic length scale. We explicitly demonstrate the connection between interface junctions and spiral patterns in these models and compute the corresponding scaling laws. This work significantly extends the results of previous studies of population dynamics and could have profound implications for the understanding of biological complexity in systems with a large number of species.
\end{abstract}

\maketitle

\section{Introduction}

Non-hierarchical interactions between individuals of different species seem to be essential to the development of the enormous biodiversity observed in nature. Rock-Paper-Scissors (RPS) type models, incorporating some of the crucial ingredients associated with the dynamics of a network of competing species, are a powerful tool in the study of complex biological systems. In its simplest version, the RPS model describes the evolution of 3 species which cyclically dominate each other \cite{Kerr2002,Reichenbach2007,Shi2010} (see also \cite{Volterra,doi:10.1021/ja01453a010} for the pioneer work by Lotka and Volterra). If the population mobility is small enough, the spatial RPS model has been shown to lead to the stable coexistence of the three species with the formation of complex spiralling patterns  \cite{Kerr2002,Reichenbach2007}.

The basic interactions behind the RPS game are Motion, Reproduction and Predation, but generalizations incorporating new interactions and further species have also been proposed in the literature \cite{PhysRevE.76.051921,Peltomaki2008,Szabo2008,Wang2010,Wang2011,Hawick2011,Hawick_2011,PhysRevE.85.051903,Avelino2012}. In \cite{Hawick2011} the standard cyclic RPS model was generalized to an arbitrary number of species and it was shown that the emerging patterns, in numerical simulations of population dynamics in a cubic grid with periodic boundary conditions, depend crucially on whether the total number of species is even or odd. While an odd number of species, if the mobility is not too large, leads to the formation of complex spiralling patterns whose characteristic length $L$ remains roughly constant in time, an even number of species is associated to the formation of partnership domains where several species coexist, whose dynamics is controlled by interactions of equal strength between species on either side of the interfaces separating adjacent domains. This has been shown \cite{Avelino2012} to lead to a scaling regime where the characteristic scale of the network $L$ grows as $L \propto t^{1/2}$, which is the typical scaling law associated with grain growth and the dynamics of soap froths \cite{Stavans1989, Glazier1992, Flyvbjerg1993, Monnereau1998, Weaire2000, Kim2006, PhysRevLett.98.145701, Avelino2010}.

In the present paper we shall consider a broad family of RPS type models, with an arbitrary number of species $N$, which may lead to the emergence of spiral patterns both for odd and even $N$. In \cite{Avelino2012} specific realizations leading to a $L \propto t^{1/2}$ scaling law have been investigated. This law was shown to accurately describe the macroscopic population dynamics of complex networks without junctions or with Y type-junctions provided that the dynamics is curvature driven, with competing species on adjacent interface domains having the same (average) strength. In the present paper we shall extend the above results to models leading to higher dimensional junctions, and investigate the connection between these interface junctions and the spiralling patterns which occur under cyclic competition between individuals from different domains. Models with an odd/even number of species shall be considered.

\section{Generalized RPS models with $Z_N$ symmetry}

Consider a model where individuals of various species and some empty sites are initially distributed randomly on a square lattice with  ${\mathcal N}$ sites. The different species are labelled by $i$ (or $j$) with $i, j= 1, ...,N$, and we make the cyclic identification $i= i+ k\, N$ where $k$ is an integer. The sum of the number of individuals of the species $i$ ($I_i$) and empty sites $I_E$ is equal to ${\mathcal N}$. At each time step a random individual (active) interacts with one of its four nearest neighbors (passive).  The unit of time $\Delta t=1$ is defined as the time necessary for  ${\mathcal N}$ interactions to occur (one generation time). The possible interactions are classified as Motion (active and passive switch their positions) 
$$
i\ \odot \to \odot\ i\,,
$$
Reproduction (active reproduces filling an empty site)
$$
i\ \otimes \to ii\,,
$$ 
or Predation (active predates the passive generating an empty site) 
$$
i\ \ (i-\alpha) \to i\ \otimes \ {\rm or}\ i\ \ (i+\alpha) \to i\ \otimes\,,
$$
where $\odot$ may be any species ($i$) or an empty site ($\otimes$), $\alpha=1,...,\alpha_{max}$ with $\alpha_{max}=N/2$ if $N$ is even or $\alpha_{max}=(N-1)/2$ if $N$ is odd). 

\begin{figure}
\centering
\includegraphics{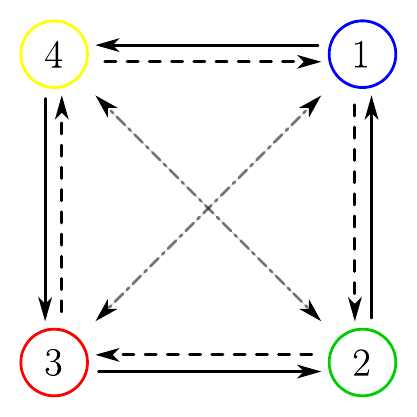}
\caption{(Color online) Scheme of all predator-prey interactions allowed in a 4-species realization of our model.}
\label{fig1}
\end{figure} 

\begin{figure}
\centering
\includegraphics{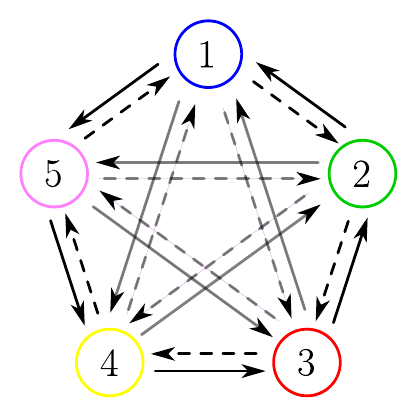}
\caption{(Color online) Scheme of all predator-prey interactions allowed in a 5-species realization of our model.}
\label{fig2}
\end{figure}

We denote the corresponding probabilities by $m_i$ (Motion), $r_i$ (Reproduction), $p_{Li\alpha}$ (left-handed Predation) and $p_{Ri\alpha}$ (right-handed Predation). In this paper we assume that $m_i=m$, $r_i=r$, $p_{Li\alpha}=p_{L\alpha}$ and $p_{R i\alpha}=p_{R\alpha}$ for all $i$, so that all the models have a $Z_N$ symmetry. We also assume that ${\mathcal N}=512^2$ and that the number densities of the various species $n_i= I_i/{\mathcal N}$ are all identical at the initial time. The number density of empty spaces, $n_E= I_E/{\mathcal N}$, is initially set to be equal to $0.1$. Figures \ref{fig1} and \ref{fig2} show a scheme of all predator-prey interactions allowed in 4- and 5-species realizations of our model, respectively. Except for the labeling of the different species, Fig. \ref{fig1} is invariant under a rotation by an angle of $2\pi/4$ thus leading to a $Z_4$ symmetry. The same is true for Fig. \ref{fig2} with $2\pi/4$ and $Z_4$ replaced by $2\pi/5$ and $Z_5$, respectively. Full arrows and dashed arrows indicate left- and right-handed predations, respectively (with probabilities $p_{L\alpha}$ and $p_{R\alpha}$ with $\alpha=1,2$). In the 4-species case  $i+2$ or $i-2$ coincide and, consequently, $p_{L2}= p_{R2}$ as indicated by the dashed-dotted arrow in Fig. \ref{fig1}. At each time step the active and passive individuals as well as the corresponding action are randomly assigned with the following probabilities
\begin{equation*}
	\begin{array}{rclcrcl}
		i\ \ (i+1) & \stackrel{p_{R1}}{\longrightarrow} & i\ \otimes, 
& \ \ \ \ & 
		i\ \ (i-1) & \stackrel{p_{L1}}{\longrightarrow} & i\ \otimes, \\ 
		i\ \ (i+2) & \stackrel{p_{R2}}{\longrightarrow} & i\ \otimes, 
& \ \ \ \ & 
		i\ \ (i-2) & \stackrel{p_{L2}}{\longrightarrow} & i\ \otimes, \\ 
		i\ \otimes & \stackrel{r}{\longrightarrow}      & i\       i, 
& \ \ \ \ & 
		i\ \odot   & \stackrel{m}{\longrightarrow}      & \odot\   i, \\ 
	\end{array}
\end{equation*}
implying that the evolution of a population network is stochastic and not fully deterministic.

\subsection{Spiral patterns}

\begin{figure}
\centering
\includegraphics*[width=4.2cm, height=4.2cm]{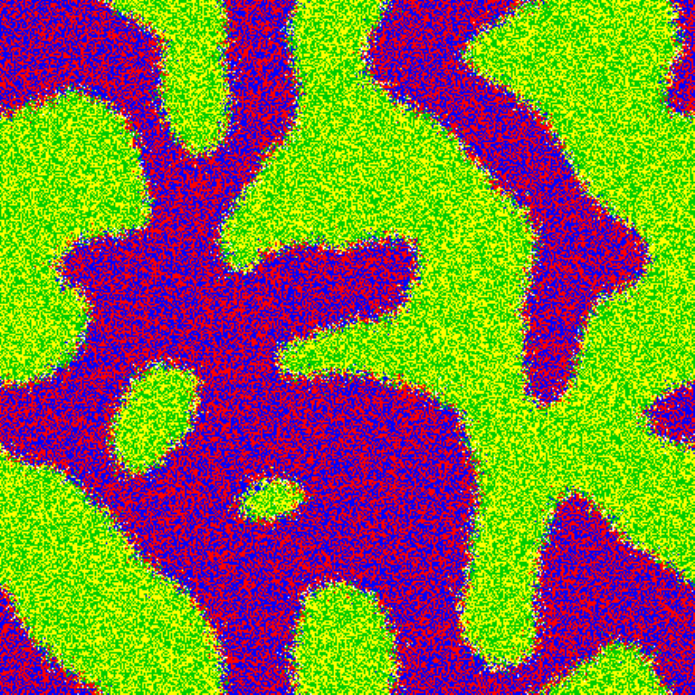}
\includegraphics*[width=4.2cm, height=4.2cm]{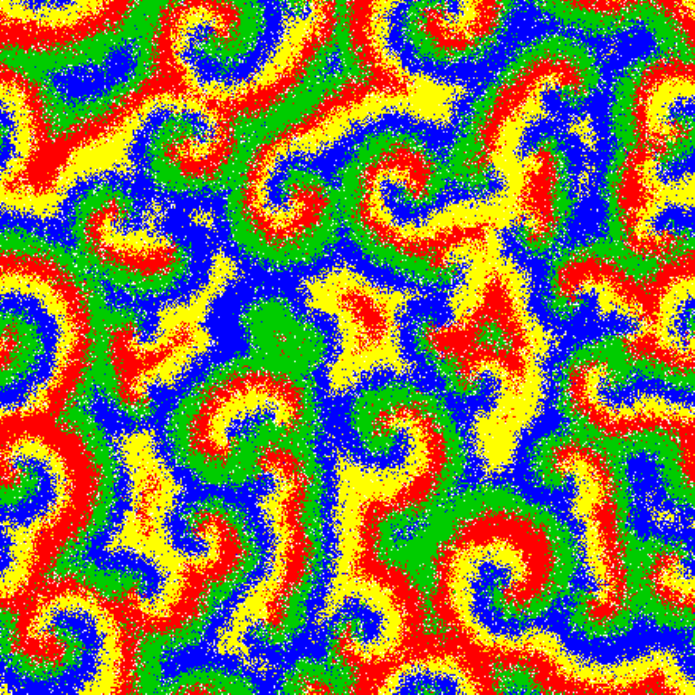}
\includegraphics*[width=4.2cm, height=4.2cm]{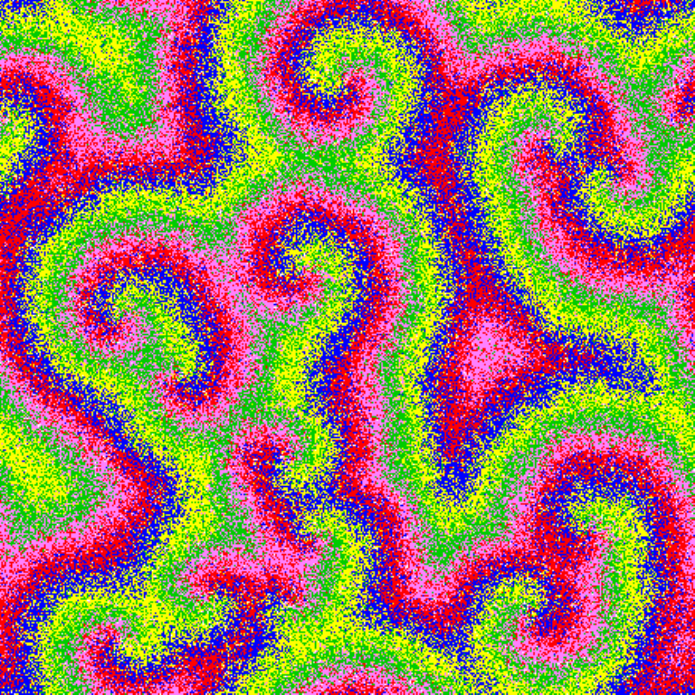}
\includegraphics*[width=4.2cm, height=4.2cm]{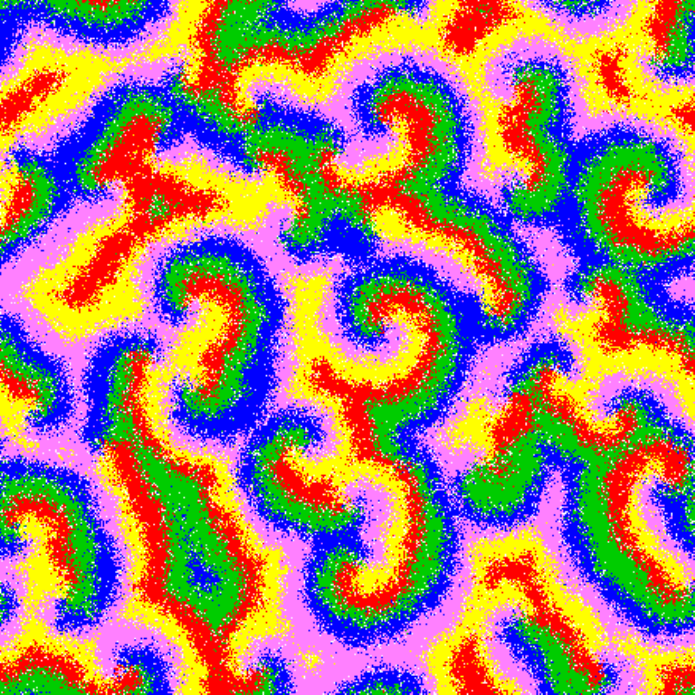}
\caption{(Color online). Snapshots of simulations of models I4 (top left), II4 (top right), I5 (bottom left) and II5 (bottom right). The snapshot of the simulation of model I4 was obtained after $1000$ generations while the others were taken after $20000$ generations.}
\label{fig3} 
\end{figure}

\begin{figure}
\centering
\includegraphics*[width=4.2cm, height=4.2cm]{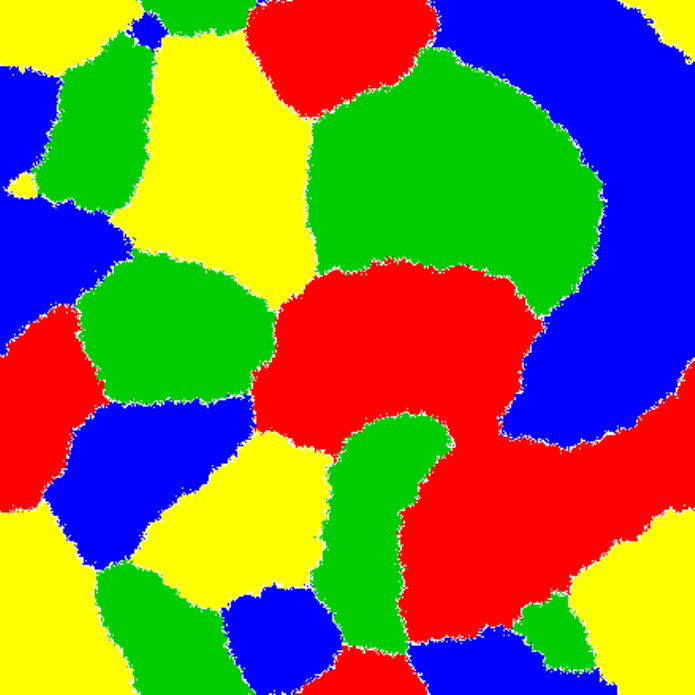}
\includegraphics*[width=4.2cm, height=4.2cm]{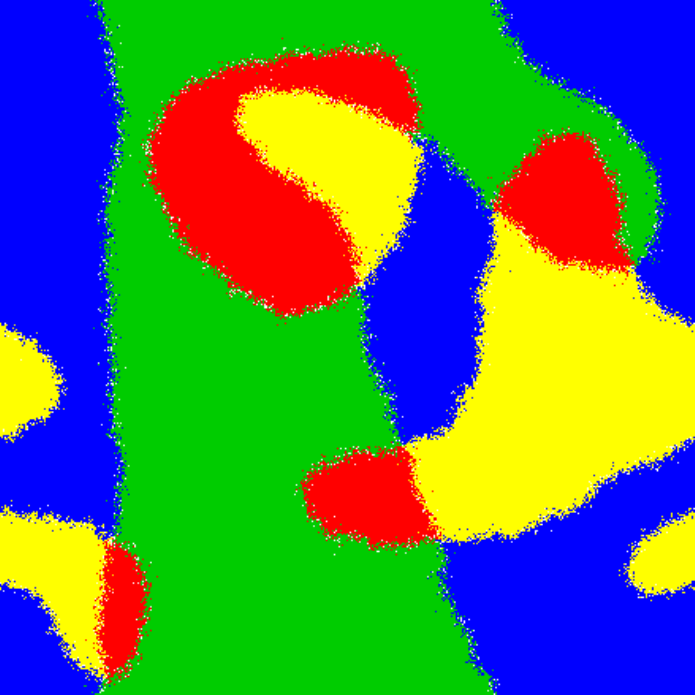}
\includegraphics*[width=4.2cm, height=4.2cm]{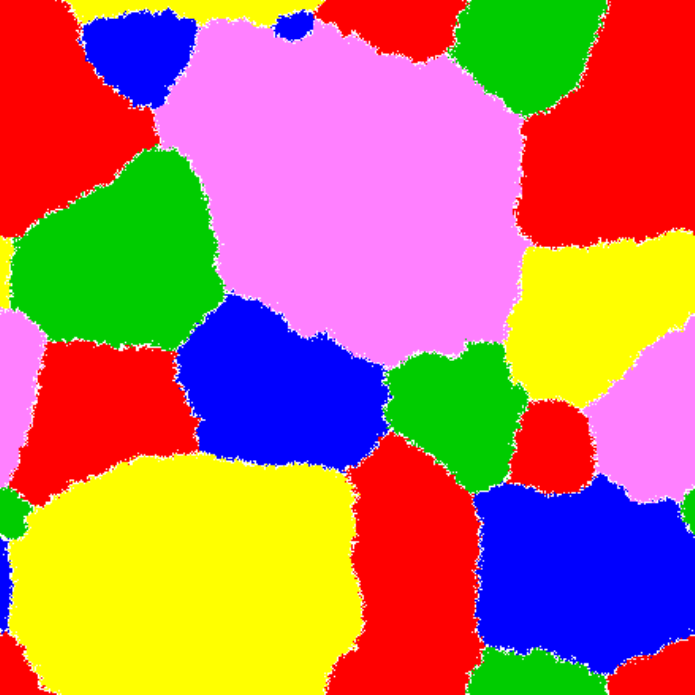}
\includegraphics*[width=4.2cm, height=4.2cm]{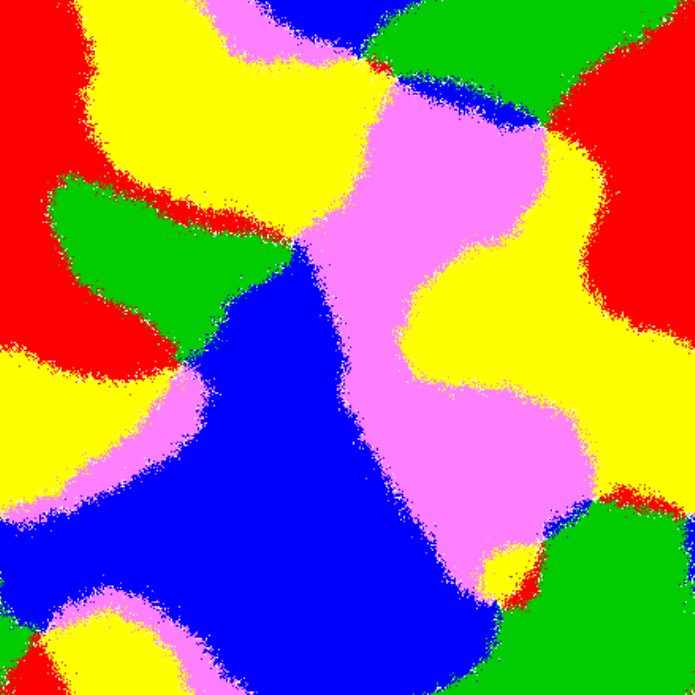}
\caption{(Color online). Snapshots of simulations of models III4 (top left), 
IV4 (top right), III5 (bottom left) and IV5 (bottom right). The snapshots were 
taken after $5000$ generations.}
\label{fig4}
\end{figure}

Consider a model with $N=4$ as a particular example of a model with an even number of species. If $p_{L1}= p$ and $p_{R1}= p_{L2}=p_{R2}= 0$ (model I4, video \cite{video1} and top left panel of Fig. \ref{fig3}), then the partnerships $\{1, 3\}$ and $\{2, 4\}$, between species which do not interact through Predation, are formed in different spatial regions. In contrast, if $p_{L1}= p_{L2}= p_{R2}= p$ and $p_{R1}= 0$ (model II4, video \cite{video2} and top right panel of Fig. \ref{fig3}) then there are no viable partnership domains and spiral patterns with 4 arms do form as a consequence of the left-handed predation between neighbouring species. For a fixed set of rules, the dynamics of the models may be very different for even and odd number of species. Consider a model with 5 species ($N=5$) with $p_{L1}= p$ and $p_{R1}=p_{L2}= p_{R2}= 0$ (model I5, video \cite{video3} and bottom left panel of Fig. \ref{fig3}). In this model the species  tend to organize themselves into spiral patterns whose $5$ arms are dominated by individuals of a species which does not interact with individuals of the dominant species of the two adjacent arms. For example, individuals of the species $1$ have as partners individuals of the species $3$ and $4$ (note that the colors blue, green, red, yellow, magenta and white in Figs. \ref{fig3} and \ref{fig4}, represent the species, 1, 2, 3, 4, 5 and empty sites, respectively). The fact that individuals of the species $i$ predate individuals of the species $i-1$ (e.g. individuals from species 4 predate individuals from species 3) is responsible for the spiral patterns in this model, even though predator and prey are never the dominant species in adjacent arms. On the other hand, if $p_{L1}= p_{L2}= p_{R2}= p$  and $p_{R1}= 0$ (model II5, video \cite{video4} and bottom right panel of Fig. \ref{fig3}) then spiral patterns with $5$ arms with a single dominant species do form, having as immediate neighbours individuals from its predator (inside) and prey (outside) species. These results can be easily extended to an arbitrary even or odd number of species by considering models with $p_{L1}= p$ and $p_{R1}=p_{L\alpha}= p_{R\alpha}= 0$ (model I) or $p_{R1}= 0$ and $p_{L1}=p_{L\alpha}= p_{R\alpha}= p$ (model II) for $\alpha \neq 1$. The even/odd asymmetry which is present in the case of model I completely disappears in the case of model II.  The results shown in Fig. \ref{fig3} were obtained with $m= 0.5$, $r= 0.35$ and $p= 0.15$. We have carried out a large number of simulations, for a wide range of parameters, and verified that the same qualitative results also hold for other choices of the parameters $m$, $r$ and $p$.

\begin{figure}
\centering
\includegraphics*[width=4.2cm, height=4.2cm]{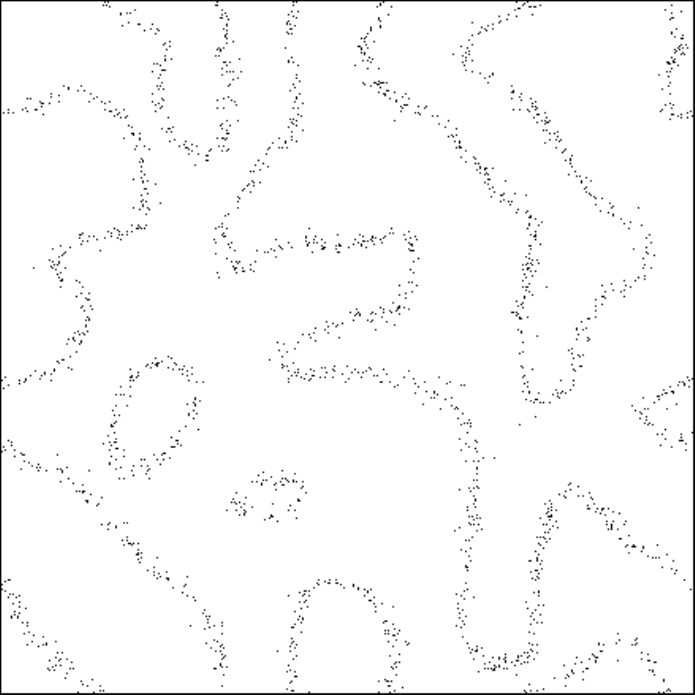}
\includegraphics*[width=4.2cm, height=4.2cm]{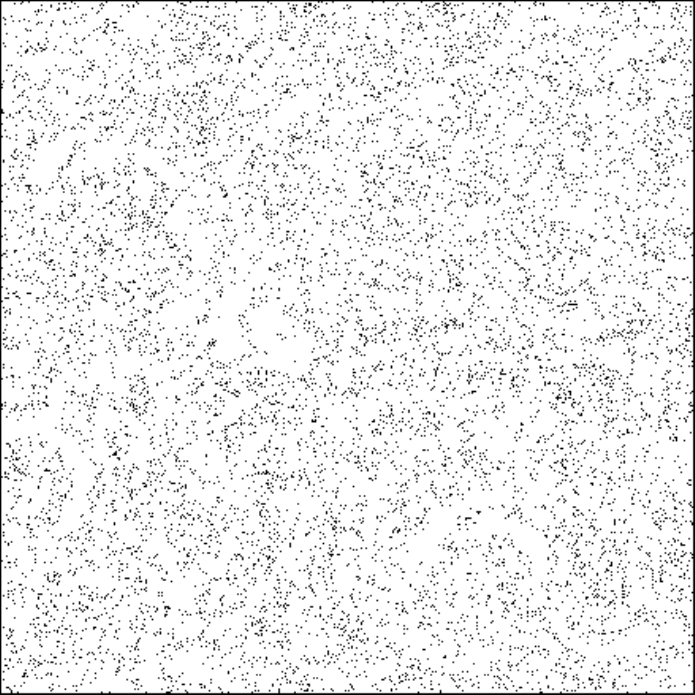}
\includegraphics*[width=4.2cm, height=4.2cm]{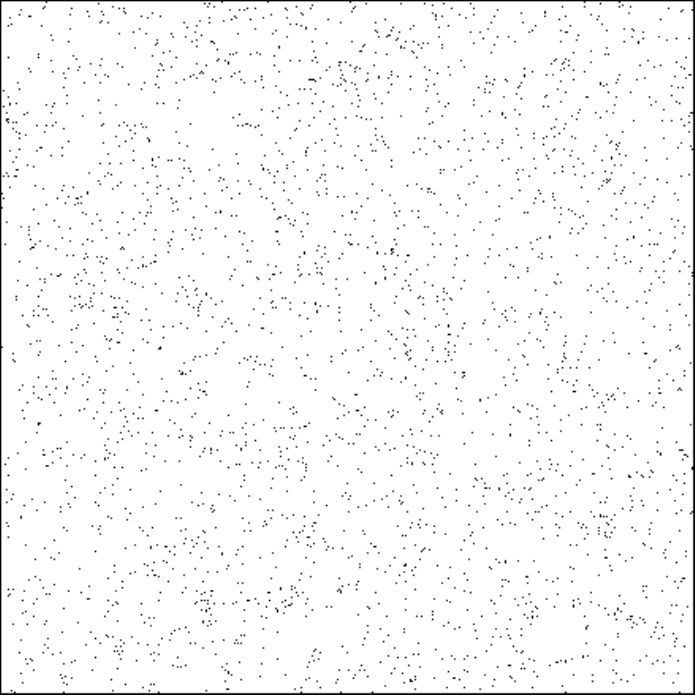}
\includegraphics*[width=4.2cm, height=4.2cm]{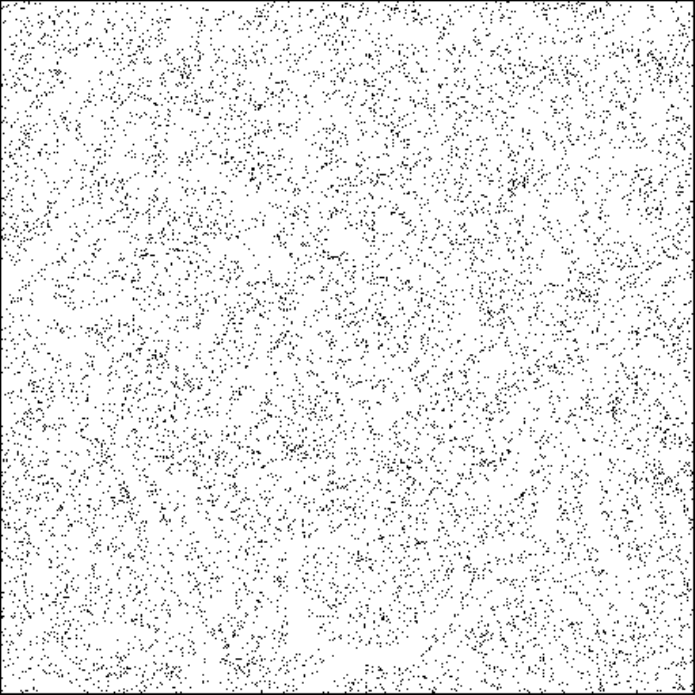}
\caption{Spatial distribution of the empty spaces of the snapshots shown in Fig.\ref{fig3} (models I4 (top left), II4 (top right), I5 (bottom left) and II5 (bottom right)).}
\label{fig5} 
\end{figure}

\subsection{Multiple junctions}

Now let us consider a model with $p_{L1}= p_{R1}= p/\kappa$ and $p_{L \alpha} = p_{R\alpha}= p$ for $\alpha \neq 1$. In this class of models the cyclic non-transitive hierarchical predator-prey rule between first neighbors does not exist and each species can hunt and be chased by any other. In this case $N$ different domain types with a single dominant species arise, separated by interfaces whose dynamics is curvature driven and controlled by interactions of identical strength between competing species. Figure \ref{fig4} shows snapshots of 4 different simulations with $N= 4$ and $\kappa=1$ (model III4, top left and video \cite{video5}), $N= 4$ and $\kappa=20$ (model IV4, top right and video \cite{video6}), $N= 5$ and $\kappa=1$ (model III5, bottom left and video \cite{video7}), $N= 5$ and $\kappa=20$ (model IV5, bottom right and video \cite{video8}). If $\kappa= 1$ then the predation rate $p_{L \alpha} = p_{R\alpha}=p$ does not depend on $\alpha$. In this case, every interface has the same effective tension, leading to the formation of an interface network with $Y$-type junctions which can be seen in the videos \cite{video5, video7} and on the left panels of Fig. \ref{fig4}. On the other hand, if $\kappa > 1$, the effective tension of the interfaces separating individuals from species $i$ and $i\pm \alpha$ is not the same for $\alpha=1$ (smaller effective tension) and  $\alpha \neq 1$ (larger effective tension). This results in the suppression of the interfaces with larger effective tensions, and may lead to the formation of stable $N$-dimensional junctions, as shown in videos \cite{video6, video8} and on the right panels of Fig. \ref{fig4}. As demonstrated in \cite{Avelino2012}, the average velocity of this type of interfaces is proportional to their curvature. As a result  the network is expected to attain a scaling regime where the characteristic length of the network $L$ obeys the scaling law $L \propto t^{\lambda}$ with $\lambda= 1/2$. Although this has been explicitly demonstrated in \cite{Avelino2012}, in the case of interface networks without junctions or with Y-type junctions, we shall provide numerical evidence that the same scaling law also applies to interface networks with higher-order junctions. All simulations shown in Fig. \ref{fig4} were performed with $m= 0.15$, $r= 0.15$ and $p= 0.70$. Although these values were found to be the most adequate for visualization purposes, we verified that many other choices of the parameters $m$, $r$ and $p$ and $\kappa$ would provide similar qualitative results.

\subsection{Empty spaces distributions}

\begin{figure}
\centering
\includegraphics*[width=4.2cm, height=4.2cm]{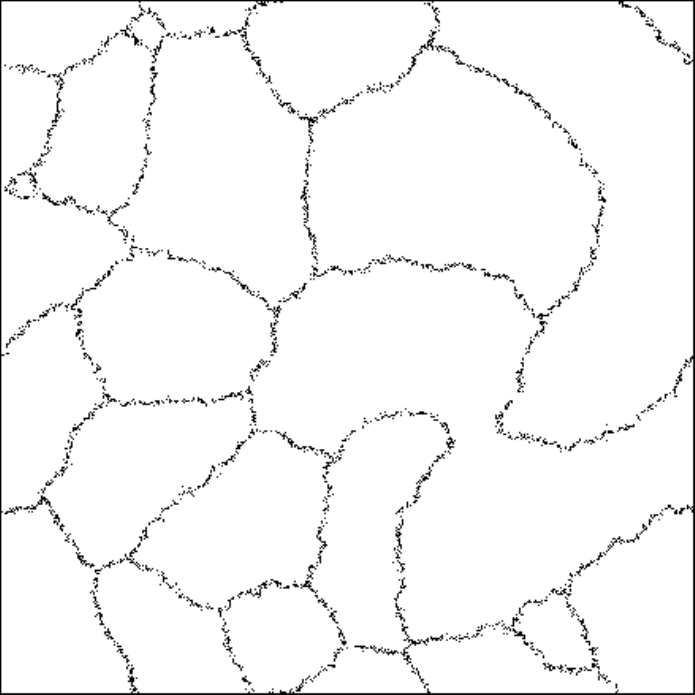}
\includegraphics*[width=4.2cm, height=4.2cm]{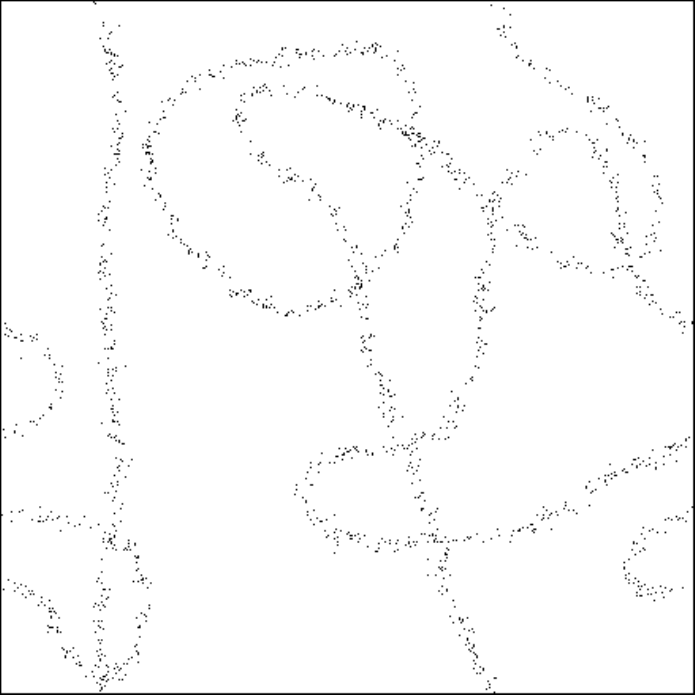}
\includegraphics*[width=4.2cm, height=4.2cm]{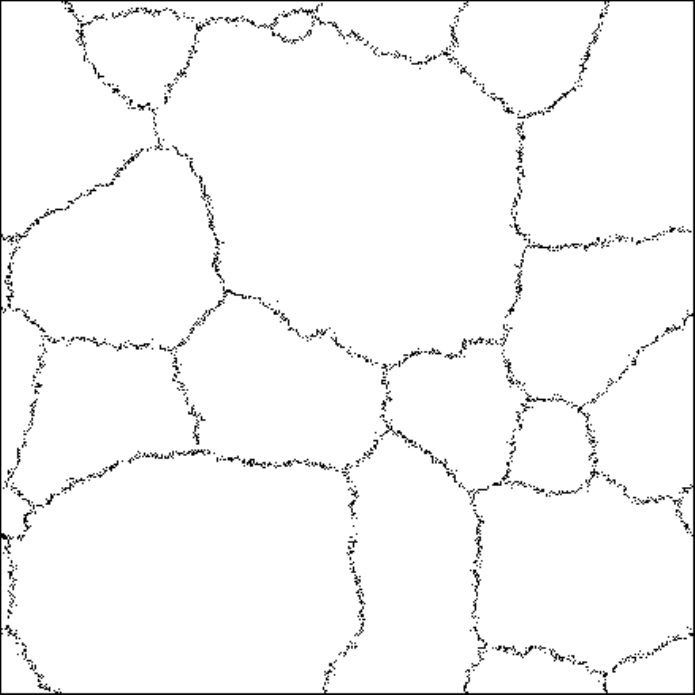}
\includegraphics*[width=4.2cm, height=4.2cm]{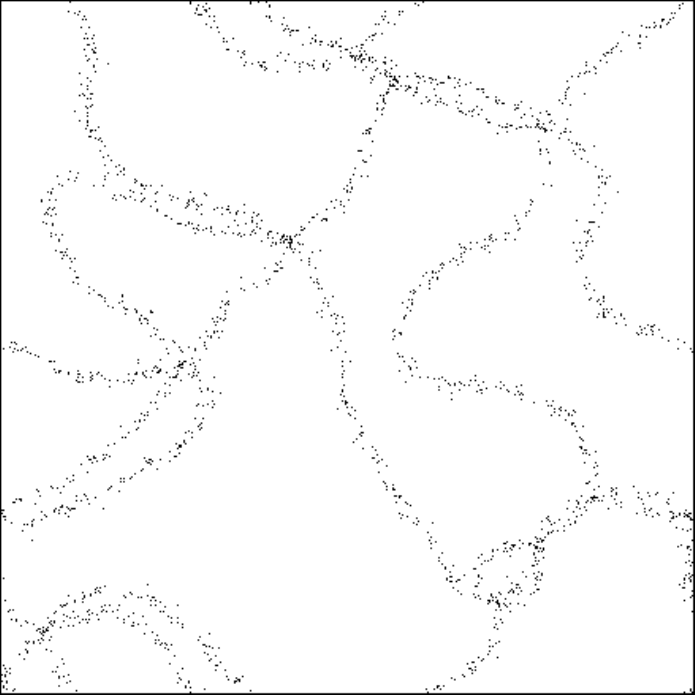}
\caption{Spatial distribution of the empty spaces of the snapshots shown in Fig. \ref{fig4} (models III4 (top left), 
IV4 (top right), III5 (bottom left) and IV5 (bottom right)).}
\label{fig6}
\end{figure}

The black dots in Figs. \ref{fig5} and \ref{fig6} represent the distribution of the empty sites of the snapshots shown in Figs. \ref{fig3} and \ref{fig4}.
These vacancies are a result of Predation between individuals from the competing species. Note that in the models I4, III and IV, they  are located at the domain borders whereas they are spread out over the entire lattice in the case of models I5 and II. Furthermore, the number density of empty spaces is much smaller in the case of model I5 than in the case of model II. This happens because in model I the species organize themselves to minimize the interactions (other than Motion), while in model II adjacent arms are dominated by individuals of interacting species. In fact, model II is the simplest generalization of the standard RPS model to an arbitrary number of species.

\subsection{Scaling laws}

\begin{figure}
\centering
\includegraphics*[width=8.0cm]{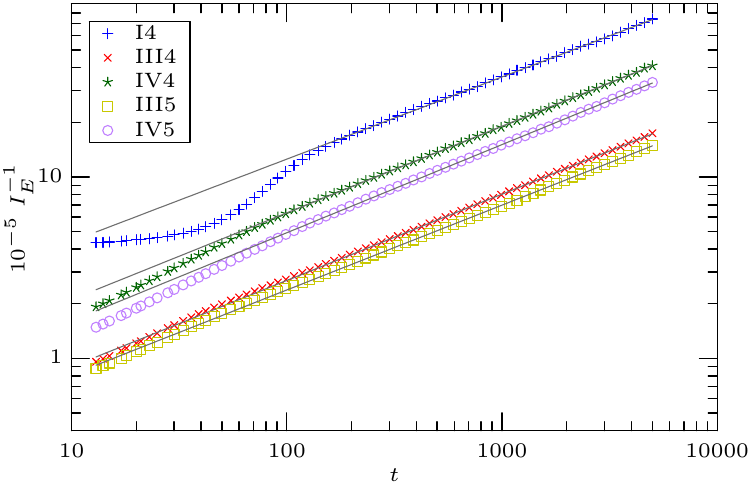}
\caption{(Color online). Time evolution of $I_E^{-1}$ for models I4, III and IV.}
\label{fig7}
\end{figure}

\begin{figure}
\centering
\includegraphics*[width=8.0cm]{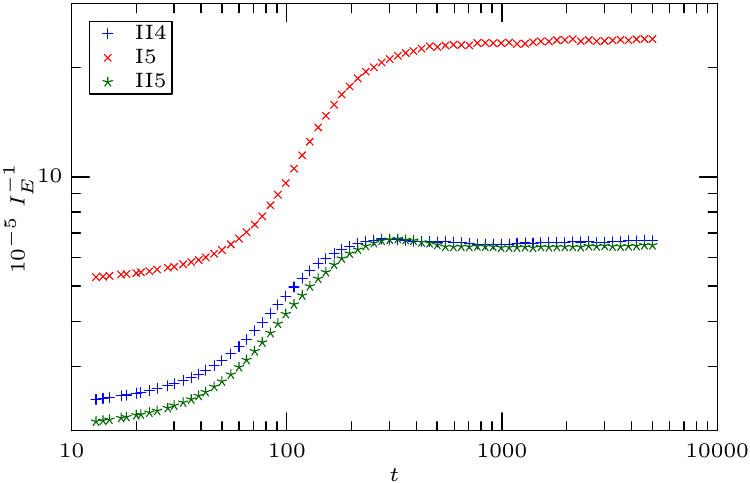}
\caption{(Color online). Time evolution of $I_E^{-1}$ for models I5 and II.}
\label{fig8}
\end{figure}

The average evolution of $I_E^{-1}$ with time $t$ (over 25 simulations with different initial conditions), in the case of models I4, III and IV,  is shown in Fig. \ref{fig7}. The scaling law $I_E^{-1} \propto t^{\lambda}$  describes quite well the late time evolution of these population networks, with $\lambda=0.460 \pm 0.063$ (model I4),  $\lambda=0.479 \pm 0.038$ (model III4),  $\lambda= 0.471 \pm 0.040$ (model III5), $\lambda= 0.483 \pm 0.055$ (model IV4) and $\lambda= 0.489 \pm 0.044$ (model IV5). These results were obtained considering only the network evolution for $t>100$. No significant dependence of the scaling exponents on the number of species and junction dimensionality have been found. In the case of models I4, III and IV empty sites appear mainly at the interfaces, whose average thickness does not change with time. Hence, the evolution with time of the characteristic length $L$ may be estimated as $L \propto I_E^{-1}$. The computed values of $\lambda$ are all very close to the $\lambda=1/2$, thus extending the validity of the above scaling law to complex interface networks with junctions of dimensionality greater than $3$. Note also that the scaling regime is achieved earlier in the case of interface networks with junctions (models III and IV), a feature also observed in the case of cosmological domain wall networks \cite{Avelino2008}. The dynamics of spiral patterns in models I5, II4 and II5 is not curvature driven. After an initial period of fast variation the number of empty sites becomes roughly constant. This behavior is shown in Fig. \ref{fig8}, where the time evolution of $I_E^{-1}$ is plotted for these models. The initial scaling regime is due to the clustering of individuals of the same species to form spiraling domains. After some time the characteristic size of the network stabilizes as the spiral patterns achieve their stationary size. 

\section{Conclusions}

In this paper we investigated the population dynamics in generalized RPS models with an arbitrary number of species, demonstrating, for the first time, that:

\begin{enumerate}

\item spiral patterns with $N$-arms may develop both for an odd and even number of species $N$;

\item interface networks with junctions of dimensionality greater than 3  may form in models with a symmetric bidirectional predation interaction between all the species, with the characteristic length scale $L$ of the network obeying the standard scaling law $L \propto t^{1/2}$ associated with grain growth and the dynamics of soap froths;

\item the simplest generalization of the standard RPS model to an arbitrary number of species, including one asymmetric $N$-cyclic predator-prey rule, leads to a population network containing $N$-armed spiral patterns with a nearly constant characteristic length scale at late times.

\end{enumerate}

This work generalizes the results of earlier investigations on the evolution of biological populations and makes new predictions for the spatial structure and dynamics of population networks whose evolution is described by generalized RPS models. Our work is expected to provide a powerful framework for the study of population dynamics and to drive the search for new signatures of population dynamics on systems with a large number of species.

\begin{acknowledgments}

We thank FCT-Portugal, CAPES, CAPES/Nanobiotec and CNPq-Brazil for financial support.

\end{acknowledgments}


\bibliography{rps2}
\end{document}